\documentclass[twoside,a4paper,11pt]{sea10}
\usepackage{graphicx}
\usepackage{hyperref}
\usepackage{amssymb}
\begin{document}
\pagenumbering{arabic}
\pagestyle{myheadings}
\thispagestyle{empty}
{\flushleft\includegraphics[width=\textwidth,bb=58 650 590 680]{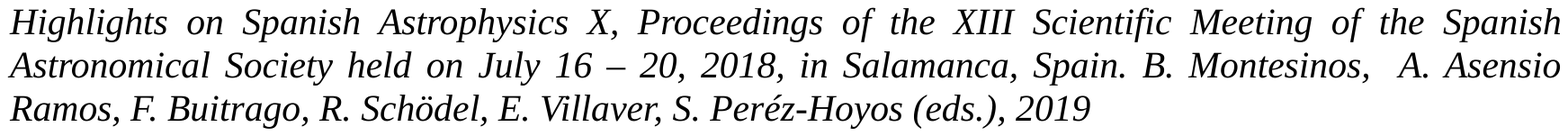}}
\vspace*{0.2cm}
\begin{flushleft}
{\bf {\LARGE
%
The tricky line of sight towards Cygnus-X: The [DB2001] CL05 embedded cluster as a pilot case
%
}\\
\vspace*{1cm}
%
D. de la Fuente$^{1}$,
C. G. Rom\'an-Z\'u\~niga$^{1}$,
E. Jim\'enez-Bail\'on$^{1}$, 
J. Alves$^{2}$, 
and 
S. Venus$^{1}$
%
}\\
\vspace*{0.5cm}
%
$^{1}$
Instituto de Astronom\'ia, Unidad Acad\'emica de Ensenada, Universidad Nacional Aut\'onoma de M\'exico, Ensenada 22860, Mexico\\
$^{2}$
Institute for Astrophysics, University of Vienna, T\"urkenschanzstrasse 17, 1180 Vienna, Austria
%
\end{flushleft}
%
\markboth{
The tricky line of sight: [DB2001] CL05
}{ 
%
D. de la Fuente et al.
%
}
\thispagestyle{empty}
\vspace*{0.4cm}
\begin{minipage}[l]{0.09\textwidth}
\ 
\end{minipage}
\begin{minipage}[r]{0.9\textwidth}
\vspace{1cm}
\section*{Abstract}{\small
%
The nearest massive star-forming complex, Cygnus-X, is widely used as a laboratory for star cluster formation and feedback processes, under the implicit assumption that all its components are located roughly at the same distance. We present a multi-wavelength study of a $15' \times 15'$ field in southern Cygnus-X, where different components involving clustered star formation are overlapped. Preliminary results indicate that the Berkeley 87 and [DB2001] CL05 clusters are actually located at very different distances, invalidating previous claims of physical interaction between them. This shows the importance of a careful treatment of extinction and distance calculations for cluster formation studies, particularly in Cygnus-X.
%
\normalsize}
\end{minipage}
%
%
%
\section{Introduction \label{intro}}

The Cygnus-X star-forming complex is usually regarded as an ideal workbench for understanding massive star formation as a whole, thanks to its proximity and richness in young star clusters and cluster-forming clouds (\cite{leduigou-knodlseder02, reipurth-schneider08}). This unique combination helps to connect the small- and large-scale processes that are involved in massive star formation, and allows to test the effects of feedback from recently formed massive clusters in detail. However, these studies have been often performed under the assumption that all the Cygnus-X components are located roughly at the same distance (e.g. \cite{knodlseder+02, motte+07, oskinova+10}), despite multiple evidence against it (e.g. \cite{pipenbrink-wendker88,uyaniker+01,maia+16}). Furthermore, \cite{uyaniker+01} claimed that interstellar clouds and OB associations are actually arranged in several layers at different distances, following the direction of the Local Galactic Arm, which is nearly perpendicular to the sky plane at Cygnus.

Unfortunately, distance estimation is particularly problematic for Cygnus-X. Due to the fact that the line of sight is roughly tangent to the Galactic rotation curve, kinematic distances (as measured by either radial velocities or proper motions) are ill-defined at $l \sim 80^\circ$, up to several kiloparsecs (\cite{ellsworthbowers+15}). Morever, \textit{Gaia} \cite{gaia+16} parallaxes are usable only to the extent that the target is unextinguished enough to be optically detected, which happens at $d \lesssim 2~\mathrm{kpc}$ in the Cygnus-X direction.

This work is focused on a small region at the southern tip of Cygnus-X, where distinct components related to recent or ongoing star formation are overlapped, namely: the well-studied young massive cluster Berkeley 87 (\cite{turner-forbes82}); the ON2 star-forming cloud hosting several masers and compact \textsc{Hii} regions (\cite{dent+88}); and an embedded cluster, [DB2001] CL05, discovered independently by \cite{comeron-torra01} and \cite{dutra-bica01}. The latter seems coincident with a clump of hard X-ray emitters that was detected by \cite{oskinova+10}. By assuming that all the components belong to a single star-forming complex, \cite{oskinova+10} interpreted these X-ray features in terms of interaction between winds from Berkeley 87 massive stars and the ON2 cloud. This scenario was subsequently challenged by the trigonometric parallax measurement of a ON2 water maser located at northern [DB2001] CL05, yielding $d = (3.83 \pm 0.13)~\mathrm{kpc}$ \cite{ando+11}, three times farther than Berkeley 87 (Cf. \cite{turner+06}).

\section{Multi-wavelength observations and extinction maps} \label{sec:obs}

\begin{figure}
\center
\includegraphics[width=6.2cm]{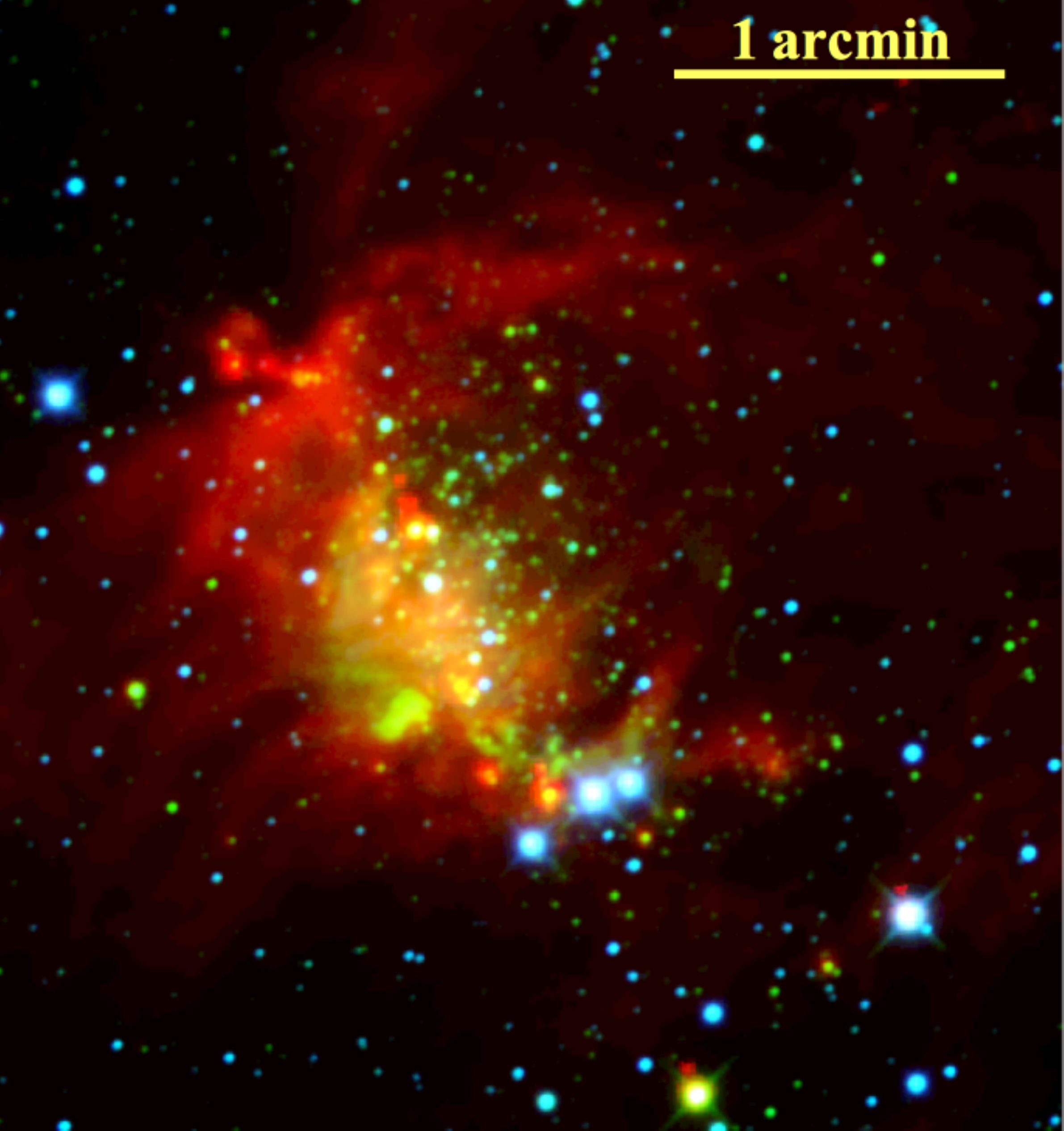}\qquad
\includegraphics[width=6.2cm]{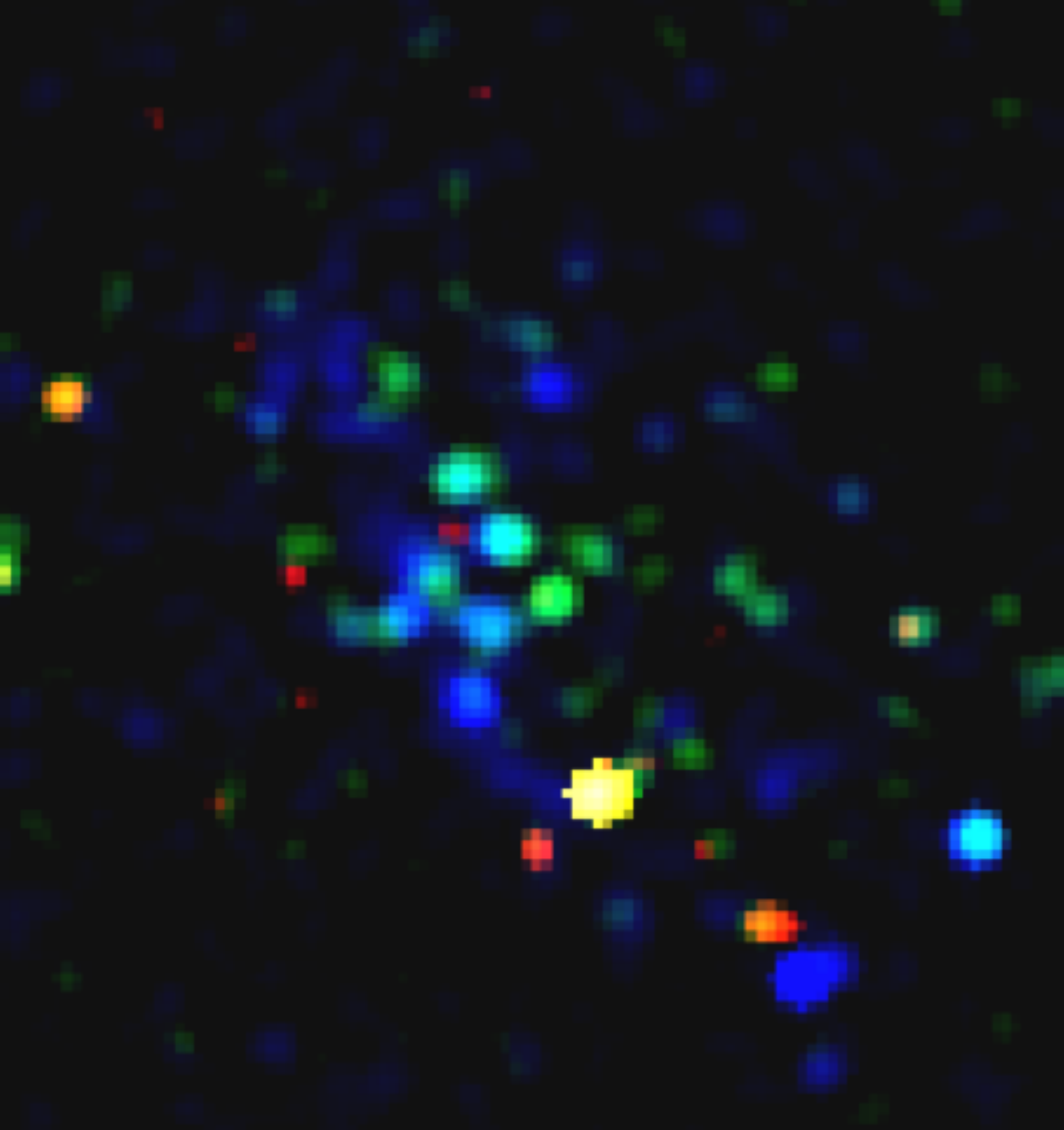}
\caption{\label{fig1} Infrared (left) and X-ray (right) RGB images covering the [DB2001] CL05 region and the overlapping part of Berkeley 87. Colors are $\mathrm{R}=[5.8]$, $\mathrm{G}=K$, $\mathrm{B}=J$ for the infrared image, and $\mathrm{R}=0.5-1.2~\mathrm{keV}$, $\mathrm{G}=1.2-2~\mathrm{keV}$, $\mathrm{B}=2-7~\mathrm{keV}$ for the X-rays. Both images cover the same coordinate ranges; north is up and east is left.
}
\end{figure}

\begin{figure}
\center
\includegraphics[width=7.4cm]{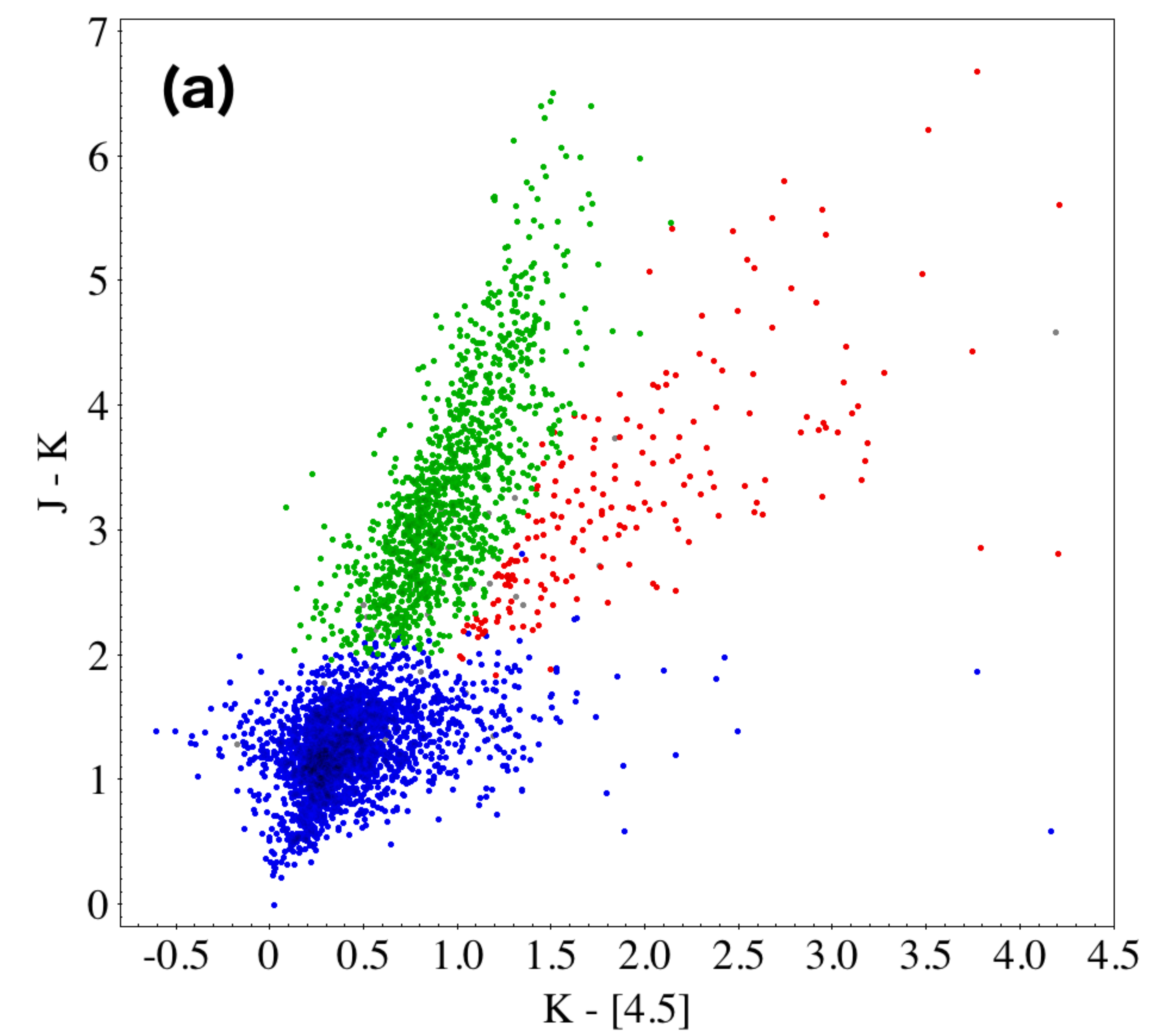}\qquad
\includegraphics[width=3.7cm]{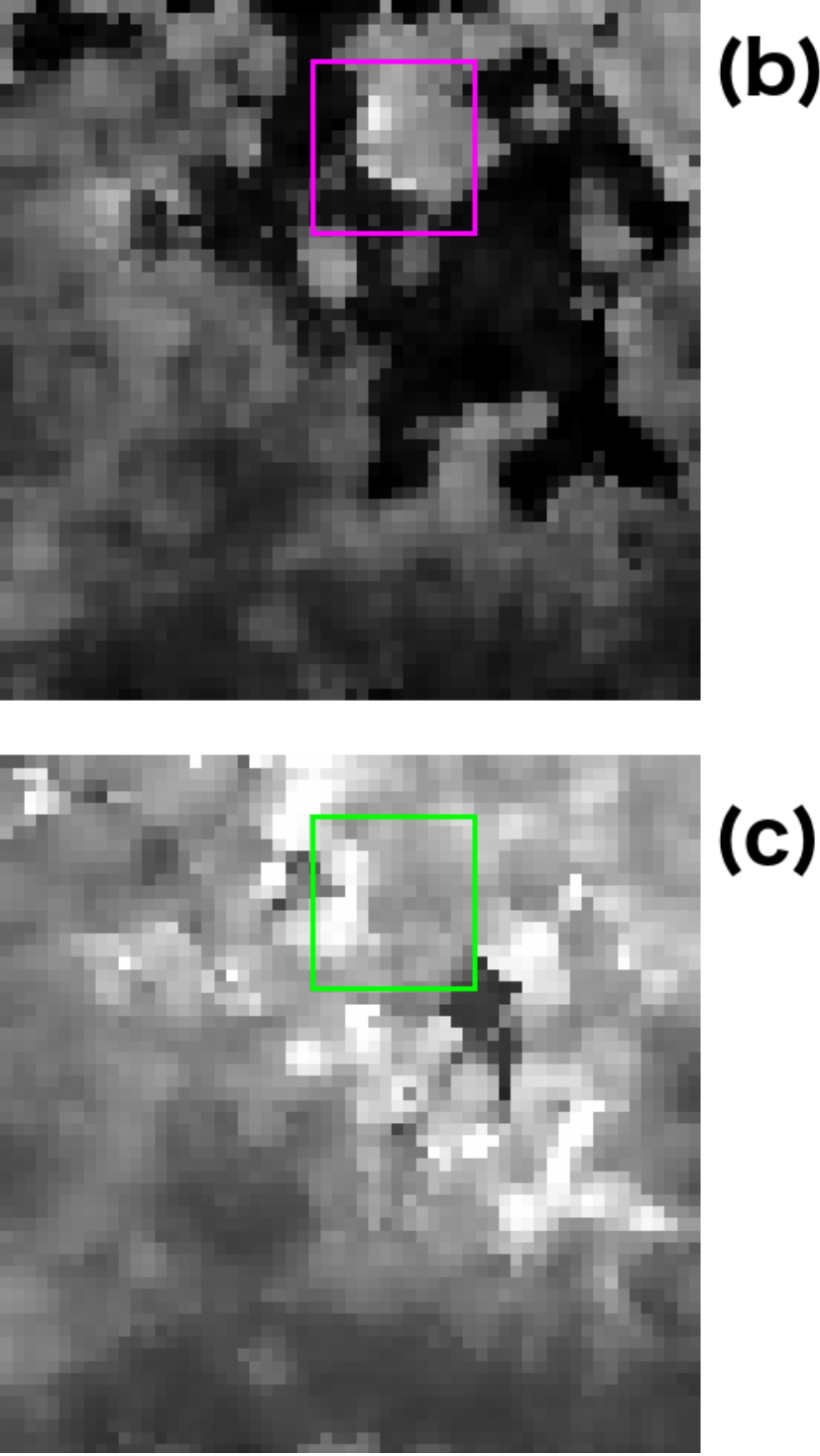}
\caption{\label{fig2} (a) Color-color diagram of point sources within the $15' \times 15'$ OMEGA2000 field, showing the low-reddening population in blue, highly extinguished stars with no intrinsic reddening in green, and highly extinguished objects with extra mid-infrared excess in red. (b) NICEST extinction map for all sources in the OMEGA2000 field. (c) Same as pannel b, but using only sources with high extinction and no intrinsic reddening (i.e. green dots in pannel a). These extinction maps are drawn in a common linear scale where black is set at $A_V=3$, and white at $A_V=25$. The colored squares enclose the region displayed in Fig. \ref{fig1}. 
}
\end{figure}

A $15' \times 15'$ field covering Berkeley 87 and [DB2001] CL05 was observed through the 3.5-meter telescope of the Calar Alto Observatory, Spain. The resulting near-infrared images and photometry are merged with those from the \href{http://irsa.ipac.caltech.edu/data/SPITZER/Cygnus-X/}{Cygnus-X \textit{Spitzer} Legacy Survey}. We also make use of archival data from the \textit{Chandra} X-ray Observatory. Fig. \ref{fig1} displays the corresponding RGB compositions in a subfield covering [DB2001] CL05, where several spectroscopically confirmed massive members of Berkeley 87 show clearly distinct colors. Specifically, the latter appear bluish in the infrared image, and orange/red (i.e. soft) in X-rays, in contrast to the souces that seem to be part of the [DB2001] CL05 overdensity. This color distinction, also seen as bimodal distributions in near-infrared colors (e.g. the two groups of data points above/below $J-K \approx 2.0$ in Fig. \ref{fig2}a), hints at a line-of-sight superposition of young stellar populations that are affected by very different amounts of foreground extinction.

Using color cuts that will be explained in detail in our forthcoming paper (de la Fuente et al., in prep.), we separate highly reddened sources with/without intrinsic reddening from the low-reddening population (Fig. \ref{fig2}a). Then, we use the NICEST method (\cite{lombardi09}) to create an extinction map using the whole near-infrared dataset. The resulting map (Fig. \ref{fig2}b) shows an extinction ``hole'' whose position and size is consistent with the Berkeley 87 cluster, but clearly contradicts the existence of interstellar clouds. On the other hand, we create another NICEST map taking into account only sources from the high-reddening group that do not show extra mid-infrared excess. The new map (Fig. \ref{fig2}c) reveals a high-extinction clumpy structure whose geometry is consistent with the ON2 star-forming clouds.

Additionally, we employ parallaxes from \textit{Gaia} DR2 \cite{gaia+18} to calculate distances for optically detectable stars (which do not include the [DB2001] CL05 cluster). A new distance for Berkeley 87, $(1669 \pm 12)~\mathrm{pc}$, is obtained as the median of spectroscopically confirmed cluster members (taking spectral types from the literature). Although somewhat higher than previous distance estimates (e.g. \cite{turner-forbes82,turner+06}), this revised value is still much lower than the aforementioned water maser distance measured by \cite{ando+11}.

\begin{figure}
\center
\includegraphics[width=12.5cm]{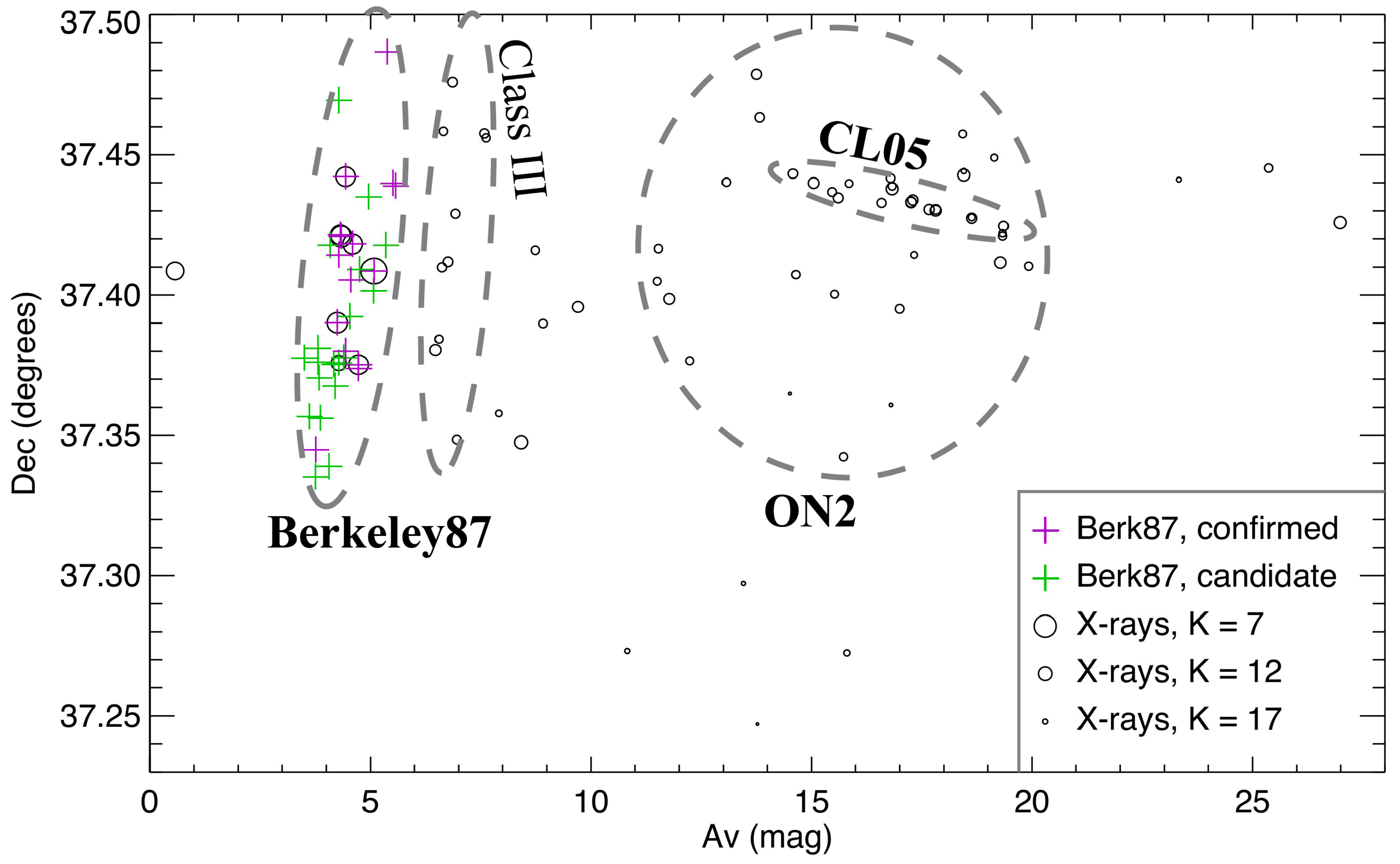}
\caption{\label{fig3} Declination vs. visual extinction for infrared counterparts of \textit{Chandra} detections within the $15' \times 15'$ OMEGA2000 field, shown as open circles whose radii are scaled with K-band magnitude. Pink and green crosses represent spectroscopically confirmed Berkeley 87 members, and further candidates listed by \cite{turner-forbes82}, respectively.
}
\end{figure}

\begin{figure}
\center
\includegraphics[width=11.9cm]{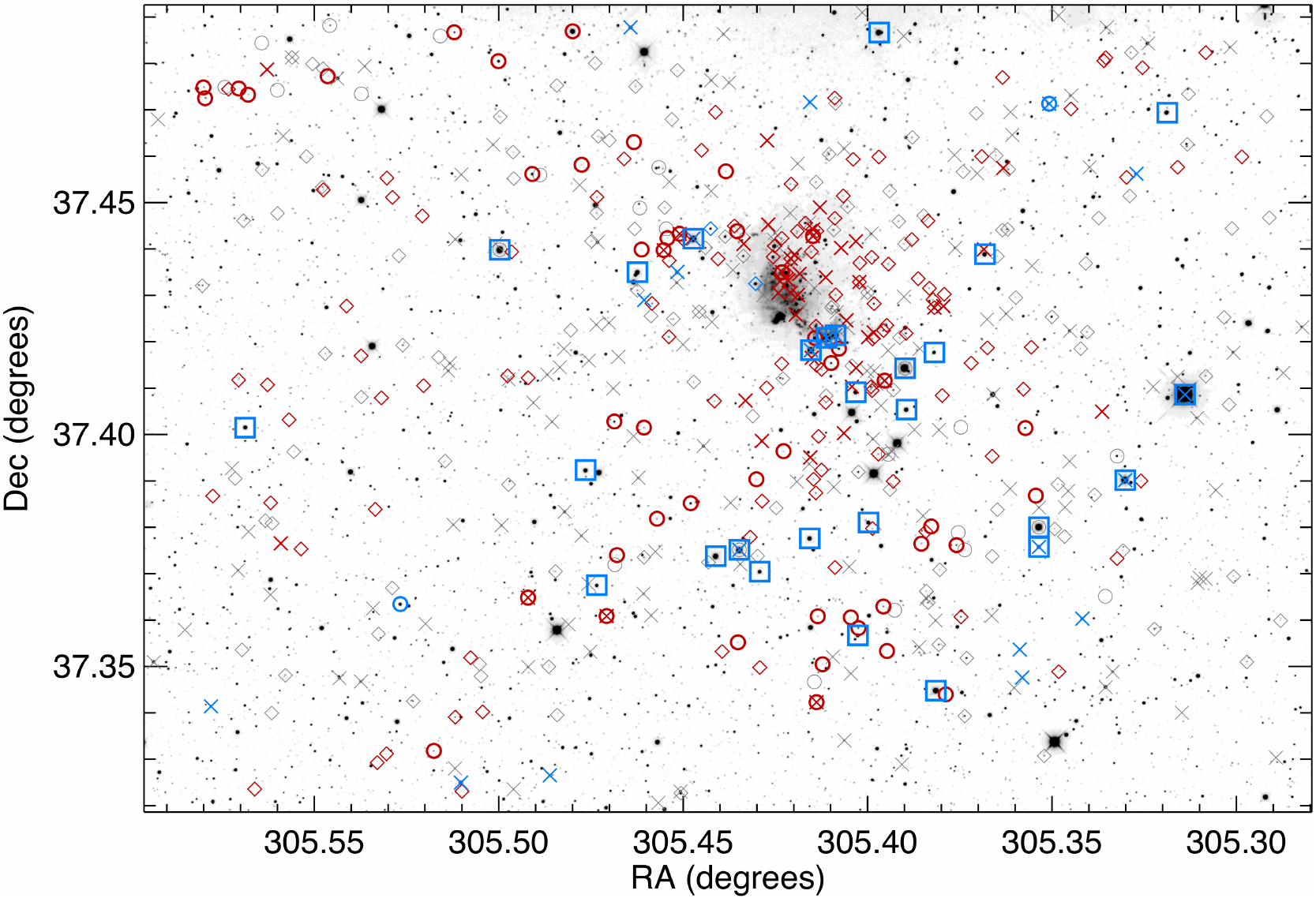}
\caption{\label{fig4} Berkeley 87 confirmed members (squares), primary (circles) and secondary (diamonds) YSO candidates, and X-ray sources (crosses) over the OMEGA2000 K-band image (excluding the less relevant southern part). Color symbols are objects assigned to the Berkeley 87 layer (blue) or the ON2 layer (red).
}
\end{figure}

\section{Layer separation}

Observational evidence presented in previous sections may indicate that the overlapping young star clusters are part of at least two physically unrelated regions, being located at very different distances. Unfortunately, distances cannot be accurately determined beyond the Berkeley 87 layer, and evidence for [DB2001] CL05 and the 3.83 kpc maser belonging to the same star-forming complex is inconclusive. Consequently, we aim at separating the Berkeley 87 and ON2 populations through extinction estimates to individual objects displaying signs of young age -- in a chronological sense. Such category includes Young Stellar Objects (YSOs) but also evolved massive stars whose ages are limited to a few Myr. In this regard, X-ray emission, which is expected to be displayed by both YSOs and hot massive stars (\cite{feigelson+07}), becomes an ideal diagnostic for inhomogeneous young populations.

Primary YSO candidates are found using the classical criteria from \cite{gutermuth+09}, and a larger amount of secondary candidates are selected through our own method for measuring intrinsic reddening (de la Fuente et al., in prep.). X-ray sources whose infrared counterparts show no intrinsic color excess are considered to be class III pre-main sequence stars, or Berkeley 87 members (note that these two options are not mutually excluding).

To provide optimal extinction estimates for as many sources as possible, several methods are combined. First, direct measurement of color excess is performed for stars of known spectral type. Second, the Rayleigh-Jeans Color Excess (RJCE) method (\cite{majewski+11}) is used in the suitable cases. Finally, the extinction map of \ref{fig2}c (whose validity is checked against sources in common with the RJCE method) is applied for stars belonging to the high-reddening group (See Sect. \ref{sec:obs}), including those that show intrinsic color excess (red dots in Fig. \ref{fig2}a). The extinction results for X-ray emitters are shown in Fig. \ref{fig3}, where several components can be clearly distinguished, and a wide gap between the Berlekey 87 and [DB2001] CL05 layers is evident. The apparent extinction shift within Berkeley 87 is attributed to class III objects experiencing residual color excesses that affect RJCE results\footnote{Note that intrinsically red YSOs in ON2 are not affected by this problem, since their extinction values are obtained through a NICEST map where such sources have been excluded from the map creation process.}.

Based on Fig. \ref{fig3}, YSO candidates with extinction values $A_V > 11$ are assigned to the ON2 layer. Moreover, any objects with signs of young ages (including spectral types) whose parallaxes are compatible with the Berkeley 87 distance are allocated in the corresponding layer. The outcome is displayed in Fig. \ref{fig4}. A vast majority of YSO candidates are located in the ON2 layer, with a strong overdensity at the position of [DB2001] CL05, while Berkeley 87 hosts only a few, mainly class III sources in the outskirts. These results are consistent with two independent clusterings of different evolutionary stage.

\section{Conclusions}

In contrast to previous claims, our preliminary results prove that star formation and X-ray emission from [DB2001] CL05 cannot be physically related to Berkeley 87 by no means, since these clusters are separated by a long distance (despite the line of sight coincidence). This case illustrates the importance of a careful treatment of extinction and distance for Galactic studies of clustered star formation, in order to avoid reaching wrong conclusions about feeback from fake neighbors.

%
%
\small  
%
\section*{Acknowledgments}   
%

D. dF acknowledges the UNAM-DGAPA postdoctoral grant. C.R.Z. and E.J.B. acknowledge support from Programa de Apoyo a Proyectos de Investigaci\'on e Innovaci\'on Tecnol\'ogica, UNAM-DGAPA, grants IN108117 and IN109217, respectively. The scientific results reported in this article are based in part on data obtained from the Chandra Data Archive. This research has made use of software provided by the Chandra X-ray Center (CXC) in the application package CIAO. This work has made use of data from the European Space Agency (ESA) mission \href{https://www.cosmos.esa.int/gaia}{\it Gaia}, processed by the {\it Gaia} Data Processing and Analysis Consortium (\href{https://www.cosmos.esa.int/web/gaia/dpac/consortium}{DPAC}).

%

%
\end{document}